\documentclass[times,twocolumn,final]{elsarticle}

\usepackage{rinp}
\usepackage{framed,multirow}

\usepackage{amssymb}
\usepackage{latexsym}

\usepackage{url}
\usepackage{xcolor}
\definecolor{newcolor}{rgb}{.8,.349,.1}

\journal{Results in Physics}

\begin{document}

\verso{Giovanni Modanese}

\begin{frontmatter}

\title{Oscillating dipole with fractional quantum source in Aharonov-Bohm electrodynamics}


\author[1]{Giovanni \snm{Modanese}\corref{cor1}}
\cortext[cor1]{Corresponding author: 
  Tel.: +39-0471-017134;  
  fax: +39-0471-017009;
\ead{Giovanni.Modanese@unibz.it}}

\address[1]{Free University of Bolzano-Bozen, P.za Universit\`a 5, Bolzano 39100, Italy}

\received{...}
\finalform{...}
\accepted{...}
\availableonline{...}
\communicated{...}

\begin{abstract}
We show, in the case of a special dipolar source, that electromagnetic fields in fractional quantum mechanics have an unexpected space dependence: propagating fields may have non-transverse components, and the distinction between near-field zone and wave zone is blurred. We employ an extension of Maxwell theory, Aharonov-Bohm electrodynamics, which is compatible with currents $j^\nu$ conserved globally but not locally; we have derived in another work the field equation $\partial_\mu F^{\mu \nu}=j^\nu+i^\nu$, where $i^\nu$ is a non-local function of $j^\nu$, called ``secondary current''. Y.\ Wei has recently proved that the probability current in fractional quantum mechanics is in general not locally conserved. 
We compute this current for a Gaussian wave packet with fractional parameter $a=3/2$ and find that in a suitable limit it can be approximated by our simplified dipolar source. Currents which are not locally conserved may  be present also in other quantum systems whose wave functions satisfy non-local equations. The combined electromagnetic effects of such sources and their secondary currents are very interesting both theoretically and for potential applications.
\end{abstract}


\end{frontmatter}



Aharonov-Bohm electrodynamics \cite{hiv} is a natural extension of Maxwell theory which allows to couple the electromagnetic field also to a current density $j^\mu$ that is not locally conserved, i.e., it is such that $\partial_\mu j^\mu \neq 0$ in some region ($\mu=0,1,2,3$). A coupling of this kind would of course be inconsistent in the Maxwell theory, since the Maxwell field equations with source are written as $\partial_\mu F^{\mu \nu}=j^\nu$ and the tensor $F^{\mu \nu}$ is antisymmetric. Aharonov-Bohm electrodynamics has only reduced gauge invariance and one additional degree of freedom, namely the scalar field $S=\partial_\mu A^\mu$ (which is a pure gauge mode in Maxwell theory).

If only locally conserved sources are present, the Aharonov-Bohm theory reduces to Maxwell theory. This happens for all classical sources and for quantum sources which obey a wave equation with locally conserved current, like the Schr\"odinger equation or Ginzburg-Landau equation. In \cite{mme} it was shown, after obtaining a covariant formulation of the Aharonov-Bohm theory and its explicit solution for $S$, that a censorship property holds: the measurable field strength $F^{\mu \nu}$ does not allow to ``see'' electromagnetically a non-conserved source $j^\nu$, because it satisfies the equation $\partial_\mu F^{\mu \nu}=j^\nu+i^\nu$, where $i^\nu$ is a non-local function of $j^\nu$, called ``secondary current'' (see eq.\ (\ref{mme2})), and $j^\nu+i^\nu$ is conserved.

Aharonov-Bohm electrodynamics can be applied in principle to systems which exhibit quantum charge anomalies \cite{cheng} or to superconducting states described by non-local equations \cite{VIG,wal}. Very recently, however, a new possible direct application has emerged. As shown by Wei \cite{wei}, the Schr\"odinger equation of fractional quantum mechanics \cite{laskin} has in general a current which is not locally conserved. This may allow particle teleportation and represents a problem for the compatibility of fractional quantum mechanics with Maxwell electrodynamics, but not for Aharonov-Bohm electrodynamics, where instead it leads to new interesting physical possibilities.

In order to illustrate these features in a paradigmatic, simplified situation, we analyze here the features of the  Aharonov-Bohm electromagnetic field generated by a non-conserved current of the form
\begin{equation}
{j_0} = \left[ {{\delta ^3}\left( {{\bf{x}} - {\bf{a}}} \right) - {\delta ^3}\left( {{\bf{x}} + {\bf{a}}} \right)} \right]{e^{i\omega t}}
; \qquad {\bf{j}} = 0.
\label{j1}
\end{equation}
This corresponds to a charge which oscillates periodically between the positions $\left( {{\bf{x}} - {\bf{a}}} \right)$ and $\left( {{\bf{x}} + {\bf{a}}} \right)$, without an intermediate current, therefore with a kind of teleportation, as allowed by fractional quantum mechanics. The source (\ref{j1}) can be seen as the limit of a suitably defined wave packet (see below).

In order to find the electromagnetic field generated by this source one must solve the Aharonov-Bohm equations, which in covariant form are
(with $\partial^2=\partial_\alpha \partial^\alpha=\partial_t^2-\nabla$)
\begin{equation}
{\partial _\mu }{F^{\mu \nu }} = {j^\nu } + {i^\nu }
\label{mme1}
\end{equation}
\begin{equation}
{i^\nu } = -{\partial ^\nu }{\partial ^{ - 2}}\left( {{\partial _\alpha }{j^\alpha }} \right)
\label{mme2}
\end{equation}

Let us define, with reference to the four-current (\ref{j1}), the quantity $\theta (t) = {\partial _\alpha }{j^\alpha } = {\partial _0}{j^0}$. Eq.\ (\ref{mme2}), which gives the secondary current, can be rewritten as $i^\nu=-\partial^\nu u_{ret}$, where $u_{ret}$ is the mathematical analogue of the retarded electric potential in Lorentz gauge of a charge $\theta(t)$, satisfying the equation $\partial^2 u_{ret}=\theta$. This allows us to compute $i^\nu$ at any point $P$ (see Fig.\ 1). Note that the potential $u_{ret}$ will not be equal to the familiar dipole potential of Maxwell theory, because the source (1) is  
different from a physical oscillating particle.

From eq.\ (\ref{mme1}) and general unicity theorems (\cite{wood}; see also \cite{cqg}), we deduce that the field $F^{\mu \nu }_R$ at any point $R$ is the Maxwell field with source $j^\nu+i^\nu$. This must include all the contributions from $i^\nu$ at each point $P$, retarded according to the distance $RP$. Note that such contributions will in general not be transversal with respect to the direction $\overrightarrow {OR} $. The distinction between near field and wave zone will also be blurred, since the secondary source $i^\nu$ extends far beyond the localized primary source $j^\nu$.

\begin{figure}[!t]
\centering
\includegraphics[scale=.37]{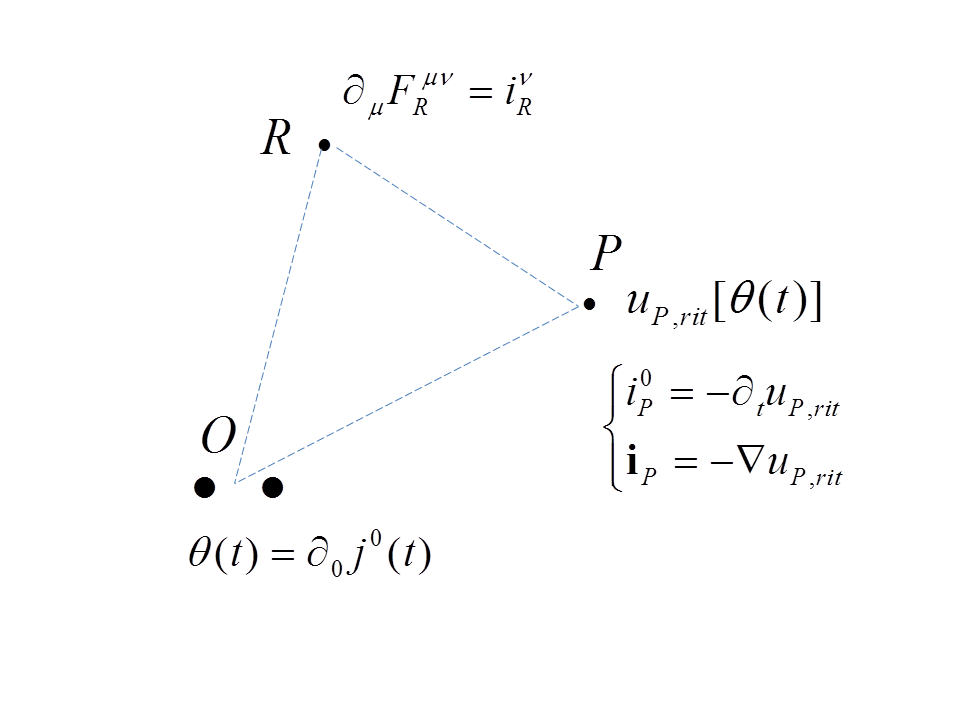}
\caption{Electromagnetic field generated by a four-current $j^\nu(t)=(j^0(t),0)$ which is not locally conserved. 
The oscillating dipolar source $\theta(t)$ in $O$ generates a secondary four-current $i^\nu$ at any point $P$ in space. $i^\nu$ is the four-gradient of $-u_{P,ret}$; $u_{P,ret}$ is mathematically equal to the retarded electric potential generated in $P$ by an equivalent charge $\theta(t)$. The observable field $F^{\mu \nu}_R$ at any point $R$ outside the dipole is found by solving the Maxwell equations with four-current $j^\nu+i^\nu$ at that point (but $j^\nu$ is zero in $R$). Therefore $F^{\mu \nu}_R$ receives ``twice retarded'' secondary contributions from any $P$ and is not generally transverse to $OR$.}
\end{figure}

Finally, let us relate our simplified Ansatz (1) to a non-conserved current in fractional quantum mechanics. According to Wei \cite{wei}, the correct probability continuity equation in fractional quantum mechanics is $\partial_t \rho+\nabla \cdot {\bf j}_a=I_a$, where $1<a\leq 2$ ($a=2$ corresponds to ordinary quantum mechanics) and the extra source term is $I_a=-iD_a\hbar^{a-1} \left[ \nabla \psi^* (-\nabla^2)^{a/2-1} \nabla \psi - c.c. \right]$. As an example, Wei computes $I_a$ for a wave function of the form $\psi=\psi_1+\psi_2$, where $\psi_1$ and $\psi_2$ are two plane waves with different energies. In order to extend this to a more realistic localized wave function, we define a Gaussian wave packet of the form
\begin{equation}
\psi=\sum_{j=1}^{n+1} \psi_j; \qquad \psi_j = e^{ -4(k_j-K)^2n_1^{-2} } e^{i\hbar k_j x} e^{-i\hbar \omega_j t},
\end{equation}
where $k_j=K_0+j$, $K=K_0+n_1$, $n_1=1+n/2$. The fractional extra source term can then be written as
\begin{equation}
\begin{array}{l}
I_a=-iD_a\hbar^{a-1} \sum_{i,j=1}^{n+1} (1-\delta_{ij})\left[ 
k_i k_j^{a-1} (\psi_i^* \psi_j - c.c.)+\right. \\
\left. \qquad + k_i^{a-1} k_j (\psi_j^* \psi_i - c.c.) \right].
\end{array}
\end{equation}

As an example, we have computed numerically $I_a$ for $a=3/2$, at $t=0$. With parameters $n=20$ (number of wavelets), $K_0=10^3$ (main wave number), $\hbar=10^{-2}$, $D_{3/2}=5\cdot 10^{-3}$ (which simply define the length and mass scale) we obtain the function in Fig. 2. By tuning the parameters we can obtain a wave packet with two peaks more localized in space and a small frequency spread; the result will be approximated well by our Ansatz (1). The full computation of $F^{\mu \nu}_R$ requires in any case a numerical solution.

\begin{figure}[!t]
\centering
\includegraphics[scale=.32]{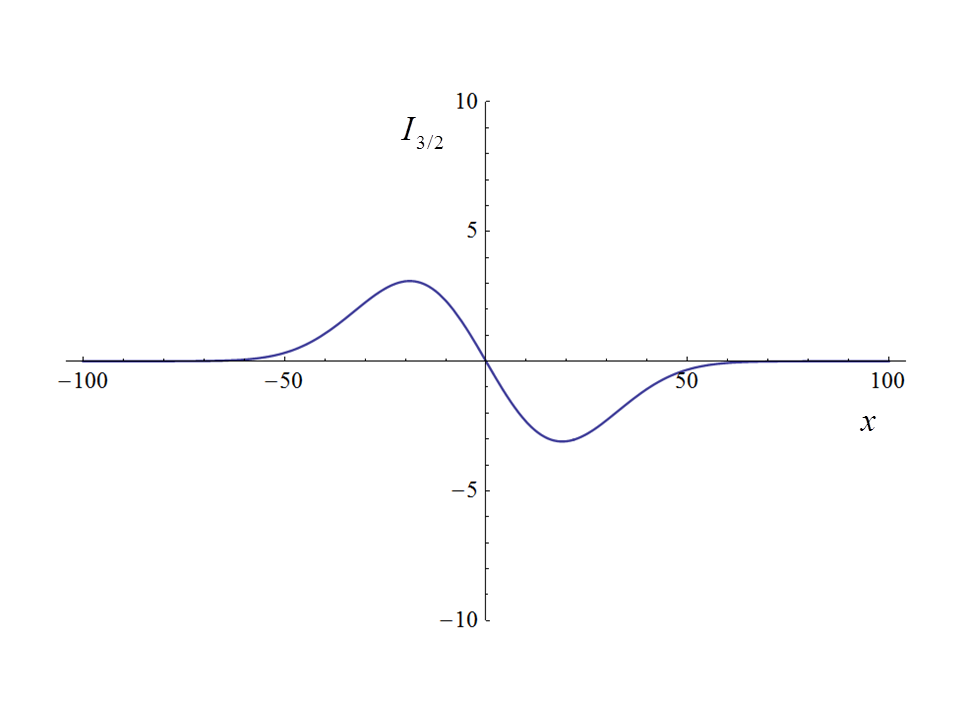}
\caption{Fractional extra-source $I_{3/2}$ for the wave packet (3).}
\end{figure}

\end{document}